\documentclass[12pt,preprint]{aastex}
\begin{document}


\title{Can Life develop in the expanded habitable zones around Red Giant Stars?}

\author{Bruno Lopez}
\affil{Observatoire de la C\^ote d'Azur} \affil{D\'epartement
GEMINI UMR 6203, BP 4229, F-06034 Nice Cedex 4, France}
\email{bruno.lopez@obs-nice.fr}
\and
\author{Jean Schneider}
\affil{Observatoire de Paris} \affil{ LUTH, 5 place Jules Janssen,
92195 Meudon Cedex, France} \email{Jean.Schneider@obspm.fr}
\and
\author{William C. Danchi}
\affil{NASA Goddard Space Flight Center} \affil{Exoplanets and
Stellar Astrophysics Laboratory, Code 667, Greenbelt, MD 20771, USA}
\email{William.C.Danchi@nasa.gov}



\shorttitle{Life in the expanded habitable zones around RGB and AGB stars?}
\shortauthors{Lopez, Schneider, \& Danchi}
\begin{abstract}

We present some new ideas about the possibility of life developing
around sub-giant and red giant stars. Our study concerns the
temporal evolution of the habitable zone. The distance between the
star and the habitable zone, as well as its width, increases with
time as a consequence of stellar evolution. The habitable zone
moves outward after the star leaves the main sequence, sweeping a
wider range of distances from the star until the star reaches the
tip of the asymptotic giant branch. Presently there is no clear
evidence as to when life actually formed on the Earth, but recent
isotopic data suggest life existed at least as early as $7 \times
10^8$ years after the Earth was formed.   Thus, if life could form and
evolve over time intervals from $5 \times 10^8$ to $10^9$ years,
then there could be habitable planets with life around red giant
stars. For a 1 M$_{\odot}$ star at the first stages of its post
main-sequence evolution, the temporal transit of the habitable
zone is estimated to be of several 10$^9$ years at 2 AU and around
10$^8$ years at 9 AU. Under these circumstances life could develop
at distances in the range 2-9 AU in the environment of 
sub-giant or giant stars and in the far distant future in the environment
of our own Solar System.  After a star completes its first ascent 
along the Red Giant Branch and the He flash takes place, 
there is an additional stable period of quiescent He core burning
during which there is another opportunity for life to develop.
For a 1 M$_{\odot}$ star there is an additional $10^9$ years with 
a stable habitable zone in the region from 7 to 22 AU.  

Space astronomy missions, such as proposed for the Terrestrial
Planet Finder (TPF) 
and Darwin, that focus on searches for
signatures of life on extrasolar planets, should also consider the
environments of sub-giants and red giant stars as potentially
interesting sites for understanding the development of life. We 
performed a preliminary evaluation of the
difficulty of interferometric observations of planets around red
giant stars compared to a main sequence star environment. We show
that pathfinder missions for TPF and Darwin, 
such as Eclipse and FKSI,
have sufficient angular resolution and sensitivity to search
for habitable planets around some of the closest evolved stars of
the sub-giant and red giant class.


\end{abstract}


\section{Introduction}

Low and intermediate mass stars with masses ranging from $0.5$ to
$\sim 8$ M$_{\odot}$
undergo dramatic changes in temperature and luminosity
during their evolution.
For example, the luminosity of the Sun, which is located on the
Main Sequence of the
Hertzprung-Russell diagram, currently is 30 percent greater
than its zero-age (initial) luminosity (Gough 1981).
Other major changes take place after the Sun evolves off the Main Sequence.
Along
the Red Giant Branch (RGB) the radius of a 1 M$_{\odot}$ star progressively
increases up to $\sim$ 10 R$_{\odot}$,
and the luminosity increases up to about
$10^{1.5}$ L$_{\odot}$ (Maeder \& Meynet 1988).
Then, after leaving the RGB a star evolves onto the
Asymptotic Giant Branch (AGB).  At this stage in its evolution,
its radius increases
up to a few hundred times R$_{\odot}$ and
the luminosity up to nearly $10^{4}$ L$_{\odot}$.
The lifetime of a 1 M$_{\odot}$ star, from the moment it leaves
the main sequence
until the end of the RGB, is of
about 2.5 $\times 10^9$ yr, which can be compared to
$\sim$ 10$\times 10^9$ yr spent on the main sequence.
The evolution of a star is accompanied
by  mass loss processes that play an important role.  During the
post-main sequence evolution of a 1 M$_{\odot}$ star nearly
0.4 M$_{\odot}$ will be lost to the interstellar medium
before it ends its life as a white dwarf (Weidemann 1987).

The ``habitable zone'' is conservatively defined as a shell
region around a star where liquid water can be found at the
surface of a planet. This zone is typically from about 0.95 AU to
1.37 AU for G stars (Kasting et al. 1993, Forget 1998). A less
conservatively defined ``habitable zone'' extends the outer edge of
the limit of the habitable zone, to as large as 2.4 AU, depending
radiative properties of $\rm CO_2$ ice clouds (Forget \& Pierrehumbert
1997, Mischna et al. 2000) . Consequently, as the temperature and
luminosity of a star changes during the course its evolution, the
radius and width of this zone will also evolve with time (see
Section~2).

The habitable zone gradually moves outward after a star leaves the
main sequence. Depending on the length of time that is necessary
for life to form, we show that it is possible that life may exist
or develop on a planet that is located within a range of 2-9 AU
from a 1 M$_{\odot}$ star, during the first stages of its post
main-sequence evolution and up to 22 AU from the star during the second
stage after the core He flash. In Section 3 this is discussed further,
considering also the effects of natural migration of
micro-organisms through meteorites as a possible trigger for the
emergence of life around evolved stars (Arrhenius 1903, O'Keefe et
al. 1996, Zubrin 2001). Micro-organisms might also transport life
from one planet to another planet during the evolution of a star.
For example, they could be transported from a planet where life is
ending to a more distant planet where favorable conditions for the
re-birth of life are encountered because of the ``sufficiently
slow'' evolution of the habitable zone.

Future space missions, such as TPF (Beichman, Woolf, \& 
Lindensmith 1999) and Darwin (Fridlund \& Gondoin 2003), 
that focus on
searches of signatures of life on extrasolar planets primarily
around main-sequence stars, should also consider the environments
of sub-giant and red giant stars. Pathfinder missions such as
EPIC (Clampin et al. 2001), Eclipse (Trauger et al. 2003),
and FKSI (Danchi et al. 2003a, 2003b) 
may also be able to observe planets around red
giant stars.  Based on a list of selected objects we evaluate the
difficulty of coronagraphic and interferometric observations of
planets around red giant stars compared to a main sequence star
environment (Section 4). In Section 5 we summarize our results and
conclude that research engaged in the field of extrasolar planets
and life should not omit searches for planets orbiting evolved
stars. Terrestrial planets in the environment of evolved stars
are sites of interest for understanding the development of life.

\section{The evolution of the habitable zone}

From a biological point of view and given the limits of our
present knowledge, the existence of life is strongly associated
with the presence of liquid water (Brack 1993). In this respect
the ``habitable zone'' is currently defined as the range of
distances from a star within which a planet may contain liquid
water at its surface. Several evaluations of the inner and outer
limits of the habitable zone have been proposed, as discussed in
the review of Forget (1998). A conservative estimate, minimizing
the width of the habitable zone (Forget 1998), assumes that around
our own Sun the habitable zone currently ranges from 0.95 to 1.67
AU. The inner limit of this conservative estimate is set by an
atmospheric water loss effect occurring in the stratosphere of a
planet, provided water vapor is injected into it from the planet's
surface (Kasting et al. 1993, Kasting 1998). The outer limit is
set by the lowest temperature at which the liquid/solid phase
change of water occurs. The estimate of this outer limit assumes
the existence of a greenhouse effect involving CO$_2$ and H$_2$O
gas (Kasting et al. 1993).  A less conservatively defined
``habitable zone'' extends the outer edge of the limit of the
habitable zone to as large as 2.4 AU, depending radiative
properties of $\rm CO_2$ ice clouds.  An optically thick condensing
$\rm CO_2$ atmosphere with particles having radii larger than 6-8
$\mu$m has a very efficient greenhouse effect that maintains the
surface temperature above the freezing point of water (Forget \&
Pierrehumbert 1997, Mischna et al. 2000).

In this paper we investigate the evolution of the transit of the
habitable zone as a consequence of stellar evolution. A convenient
way of estimating the inner and outer limit of the habitable zone
at different stages of stellar evolution is to assume that the
planet behaves as a gray-body with albedo, $A$, and with perfect
heat conductivity (implying the temperature is uniform over the
planet's surface). Its radiative equilibrium temperature, $T_{p}$,
is then determined by~:

\begin{equation}
T_{p}=[(1 - A)L_{\star}/(16 \pi \sigma d^2)]^{0.25}
\end{equation}

where $L_{\star}$ is the luminosity of the star, $\sigma$ is the
Stefan-Boltzmann constant, and $d$ is the distance from the center
of the star to the planet. Typically a value of approximately 0.2
is chosen for the albedo to be representative of an earth-like
planet.  For these calculations we determine the temperature at
the limits of the habitable zone, and from this we calculate the
inner and outer distances for a given star.  If the albedo is
wavelength independent, then terms due to the albedo cancel out of
the calculation.

In the following calculations we use both the conservative limits
of the habitable zone and the less conservative limit around our
Sun (Forget 1998, Forget and Pierrehumbert 1997, Mischna et al.
2000). For the conservative limit, we assume an effective
equilibrium temperature of 269 K and 203 K, which for the Sun
gives inner and outer limits of 0.95 and 1.37 AU, respectively.
For the less conservative limit, the equilibrium temperature is
169 K, so that the outer limit is extended to 2.4 AU. The three
equilibrium temperature extrema, defined as the inner and the 2 possible outer
limits, can be used to derive the approximate location and width
of the habitable zone around a star at different stages of its
evolution.

Models describing the stellar evolution are proposed by Maeder and
Meynet (1988). For intermediate mass stars having masses from 2 to
5 M$_{\odot}$ the evolution is followed up to the end of the early
asymptotic giant branch. For stars with masses less than 1.7
M$_{\odot}$, the evolution is calculated up the He-flash occuring
at the end of the RGB part of the HR diagram (Iben 1967).  Maeder
and Meynet's (1988) models are used in the following figures to
determine the effect of stellar evolution upon the habitable zone,
i.e., its outward transit and increase in width.

Figure 1(a) displays the time evolution of the inner and outer
limits defining the habitable zone assuming the initial mass of
the star is 1 M$_{\odot}$. The solid curve represents the
evolution of the inner limit of the habitable zone, while the
dashed curve depicts the conservative outer limit, and the dotted
curve represents the less conservative outer limit. The distance
between the center of the star and the habitable zone increases
progressively with time, from 1 AU to tens of AU (upper limit
represented in the Figures). Note that it is easy to show from Eq.
(1) that this distance is proportional to the square root of the
stellar luminosity, which greatly increases as the star ascends
the Red Giant and Asymptotic Giant Branches of the HR diagram.
The width of the habitable zone also increases with time
and is also proportional to the square root of the stellar
luminosity. Thus the width is directly proportional to the
distance between the star and the habitable zone.

Therefore, during the course of stellar evolution, the habitable
zone is a shell sweeping progressively outward over a wide range
of distances from the star as can be seen in Figure 1(c).  The
dashed curve represents the duration of the habitable zone for the
conservative assumptions for the inner and outer limits of the
habitable zone.  The dotted curve assumes the same inner limit of
the habitable zone but the less conservative outer limit is used.
The duration of the transit during which the habitable zone passes
over a planet located at 1 AU from a star is of the order of
10$^9$ years. 
Figure 1(b) displays the evolution of the stellar radius (in units
of Solar radii) during the time period shown in Figure 1(a).  
Immediately after the star leaves the main sequence, the habitable
zone progressively moves to 2 AU (Figure 1(c)). 
The duration of the transit at
this location is approximately 10$^9$ years. A bump/plateau is observed in
the curve up to 9 AU (for the conservative limits) and up to 13 AU (for the
less conservative limits), where 
the duration of habitable conditions
lasts from a few to several 10$^8$ years. At 10 AU the duration is
smaller, around 10$^8$ years. At 15 AU from the star, the duration
of habitable conditions lasts more than 10$^7$ years, and at the
largest distances the duration gradually decreases. Thus the
evolution of the habitable zone at distances up to 25 AU has been
considered. During this time period, a 1 M$_{\odot}$ star has not
yet reached the AGB.

The bump/plateau observed up to 9-13 AU increases 
the duration of habitability conditions.
This bump results from a local maximum in luminosity followed
by a local minimum along the RGB related to the first dredge-up.

For a 1.5 M$_{\odot}$ star, the evolution of the habitable zone
is much more rapid since the star itself evolves faster than a
1 M$_{\odot}$ star. Plots of the inner and outer limits of
the habitable zone, the radius of the star, and the 
duration of the habitable zone for this case is shown in
Figs. 2(a), 2(b), and 2(c), respectively.
When the star is on the main sequence,
the transit of habitable zone, at 3
AU, lasts more than 10$^9$ years. At 5 AU the transit may
last from 10$^8$ to several 10$^8$ years (depending on the limits assumed for
the habitable zone). Interestingly, even though the star
evolves rapidly, the duration of the passage of the habitable zone
at distances less that 15 AU from the star remains greater than 10$^7$ years.
Bumps are observed in the curve since the variations (just after
the main sequence and along the RGB ascent in the HR diagram)
of the effective temperature and luminosity of the star is non-monotonic.

For a 2 M$_{\odot}$ star (Figs. 3(a)-3(c)) the habitable zone is at $\sim$ 5 AU
during the main sequence stage. After the star leaves the main sequence
the transit of the habitable zone may last more than 10$^8$ years for
distances less than 10-15 AU.
It is interesting to note that the habitable zone 'jumps' at about 1.7 Gyrs
over a period lasting less than 25 million years (see
Figure 3(a)).

In this paper we have calculated the duration of the habitable zone for
different stellar masses, but we have not considered the {\em actual}
mass distribution of red giants in our own Solar neighborhood.  Clearly
the longest lived habitable zones are those for the low mass red giant
stars that evolve most slowly off the main sequence, as is evident from
the present discussion and Figs. 1-3, as well as Table 1, described below.

Although individual masses of red giant stars are not well known in general, 
and have not been directly determined except in a few cases, Neckel (1975)
has shown theoretically that the peak of the red giant mass distribution occurs
at progenitor masses of about 1.1 M$_{\odot}$. An empirical estimate
of red giant masses determined by Scalo, Dominy, and Pumphrey (1978) is 
in agreement with the theoretical estimate, with values 0.8 -- 1.2 M$_{\odot}$.  This is fortunate 
since stars selected in a magnitude-limited or distance-limited survey
will therefore mostly likely be low mass stars that have long habitable
zone lifetimes similar to those calculated in Fig. 1 in this paper.

Also, it is important to note that the calculations in this paper are 
focused on
the evolution of the habitable zone from the time period after the 
star evolves off the main sequence up the first ascent along
the Red Giant Branch during which the star is burning hydrogen in
a shell around a growing He core, i.e., only to the point of the helium
flash.  After the helium flash, there is an additional long-lived phase
of quiescent core helium burning with nearly constant luminosity.  

Scalo and Miller (1979) developed
very simple interpolation formulae for the lifetime of a star at 
its various stages of evolution:  main sequence, subgiant, first red giant,
and first and second helium burning phases, as a function of initial 
stellar mass.  For completeness we display the lifetimes computed from
their formulae for stellar masses of 1.0, 1.5, and 2.0 M$_{\odot}$, which
we display in Table 1.  For a 1.0 M$_{\odot}$ star, core helium
burning provides a nearly constant luminosity phase, adding an additional
$10^9$ years to the duration of the habitable zone for low mass stars, 
with substantial increases to the lifetime of the habitable zone for 
higher mass stars as well.  The luminosity during the core helium burning phase
is approximately $2.3 \times 10^2$ L$_{\odot}$,  $9.3 \times 10^2$ L$_{\odot}$,
and $2.3 \times 10^2$ L$_{\odot}$, for 1.0, 1.5, and 2.0 M$_{\odot}$ 
stars, respectively (Kozlowski \& Paczynski, 1975). Thus 
an additional period of habitability exists for the 
region between 7 and 22 AU for a 1 M$_{\odot}$ star during core helium 
burning.  

The He core flash that precedes this phase is not an explosive event nor
does it affect the star's external luminosity very much.  This was discussed
in the pioneering papers by Schwarzschild \& Harm (1962, 1964), and subsequently
analyzed in more detail by Mengel \& Gross (1975), Paczynski \& Tremaine (1977), Paczynski (1979), and by Deupree (1996).  
The main consequence to the stellar luminosity after the He core flash
is an initial {\em decrease} in luminosity for about $10^3$ to $10^4$ years,
followed by an increase to approximately the same value as before the
flash, i.e., near the peak luminosity at the tip of the first ascent 
along the Red Giant Branch (Despain, 1981).
Over a period of $10^6$ to $10^7$ years
the star gradually becomes less luminous, and it settles down to a 
state of nearly constant luminosity of about $10^2$ L$_{\odot}$ 
(Despain, 1981; Chiosi, Bertelli, and Bressan, 1992)  
Statistically the result is a ``clump" of stars of this luminosity
close to the RGB.  The constancy of luminosity for this ``clump'' of
stars, which is about 
a factor of 10 lower than at the tip of the RGB, is a result of the
fact that stars of different initial masses converge toward a 
common value of the He core mass (Chiosi, Bertelli, \& Bressan, 1992).

For a 1 M$_{\odot}$ star the duration of the rise in luminosity
as a star evolves upwards along the RGB is relatively rapid, lasting
$\sim$ $2 \times 10^7$ years as can be seen in Fig. 1.  During this period
of time the habitable zone moves rapidly outward, such that the 
inner radius is $>$ 20 AU from the star.  Once the star settles into
the core He 
burning phase, the region between 7 and 22 AU becomes habitable 
again.  Thus there is a ``fresh start" for life in this phase 
of stellar evolution.  

We summarize our analysis of the evolution of the habitable zone
by noting that there are two opportunities for life in the 
extended habitable zones around evolved stars.  Initially 
life has an opportunity as the star slowly evolves 
off the main sequence.  After the first ascent along the
RGB and the He core flash, there is a second chance during the core He
burning phase.  This latter phase is particularly interesting 
because of the nearly constant luminosity of the star and the 
long duration under these conditions.   

\section{The possibility of the existence of life around Red Giant Stars}

The Earth is known to have formed about 4.55 Gyr ago. Substantial
evidence exists that life on Earth occurred at least as far as
3.85 Gyr ago rests on the basis of isotopic ratios of carbon
residues within grains of apatite (Mojzsis et al. 1997, Holland
1997).  The oldest purported fossils of microorganisms have an age
of 3.50 Gyr and may represent photosynthetic cyanobacteria were
reported by Schopf (1993, 1994).  The findings and interpretation
of this work was recently debated (Brasier et al. 2002, Dalton
2002). Rocks older than 3.60 Gyr are all metamorphosed at a high
grade and most are strongly deformed, thus morphological fossils
if originally present are unlikely to be preserved (Rosing 1999).

Based on these considerations, it is possible for life to have existed
at least as long ago as 3.85 Gyr, but this is uncertain because of
the imprecise knowledge of the ages of the rocks, and ambiguous
interpretations of the isotopic $\rm {^{12}C/^{13}C}$ ratios.  Thus, for our
discussion, we consider three time scales for life to form, a
conservative estimate of 1 Gyr, and a less conservative one of 0.5
Gyr, and an optimistic one of 0.1 Gyr.  
Recent analyses of isotopic $^{13}$C ratios on micron-scale mineralized
tubes in Archean pillow lavas by Furnes et al. (2004) suggest 
that microbial life colonized these oceanic volcanic rocks soon
after they were formed approximately 3.5 Gyr ago.  This provides
some further support for the optimistic time scales of a few hundred
million years or less for life formation.

Dating of terrestrial fossils reveals that life had evolved on
Earth by the end of the period of heavy bombardment. Evidence for
the bombardment comes from theoretical studies of planetary
formation and directly from lunar cratering. Heavy bombardment may
have helped set the stage for the terrestrial origin of life by
delivering key biogenic elements (H, C, N) to the Earth's surface
(Chyba et al. 1994). For its first 800 Myr, the Earth's surface
would have been subjected to frequent impacts, possibly delaying
the development and spread of life (Maher and Stevenson 1988).
Sleep et al. (2001) estimate that the time period for the Earth to cool
off after heavy bombardment to a temperature at which life could
form is surprisingly short, in the range of $10^5$ to $10^7$ years.

Due to the rapid evolution of a star once it has left
the main sequence, the transit time of the habitable zone shortens.
For an evolved 1 M$_{\odot}$ star, the transit of the habitable
zone crossing a planet at 2 AU of distance lasts slightly more than 10$^9$ years,
a time scale clearly larger than the time which has been required for the emergence
of life on Earth.

At about 5 AU of distance, the duration of habitable conditions
lasts from 10$^8$ years to several 10$^8$ years , which 
is about one order of magnitude shorter
than the time estimated for occurrence of life on Earth. Even
if the formation of life takes longer than 10$^8$ years, it is
still possible for pre-existing life (in the form known on Earth)
to adapt itself to a new planet containing liquid water.

Micro-organisms could be transported from a
planet where life is ending to one where favorable conditions for
its re-birth are encountered due to the transit of the habitable zone.
In this sense the transportation of microorganisms through
meteorites can be a trigger for life development around evolved stars.
The possible migration of life through meteorites has been discussed by
Arrhenius (1903), O'Keefe et al. (1996) and Zubrin (2001), while
the possibility of micro-organisms transporting life
in a planetary system either
from planet to planet or between different planetary systems has also
been studied (Mileikowsky et al. 2000, Mastrapa et al. 2001).
The transportation of material between planetary systems has recently
gained support from the possible discovery of extrasolar micro-meteors
(Meisel et al. 2002).
Of course, the time scales are longer because
the distances are larger. Meteorites or comets may have been expelled
from our Solar System up to 3.5 billion years ago (when life first
appeared on Earth) with biomolecules or even primitive microorganisms on them.
After a journey of a billion years or more in interstellar space, they
may have been captured by nearby planetary systems
(up to a few tens of light years)
and may have seeded other planets with suitable conditions for the development of life.
Mathematical simulations have shown that the probability for expulsion of
a meteorite or a comet from a planetary system is not negligible 
(Gladman et al 1996).
Current theoretical debate centers on the probability of capture by an
extrasolar planet.  However, recent calculations by Melosh (2003) suggest
that the probability of interstellar transfer of meteoric material 
between planets on two different stars in the Solar neighborhood (for 
single stars) is extremely small.  However in the case of red giant 
environments, the probability of capture of meteoritic material ejected
from a planetary body is reasonably large (Melosh, 2003).  Using this mechanism, 
life can be transported from one planet to another during the expansion
of the habitable zone.

It has been demonstrated both by
laboratory and space experiments that biomolecules and even
micro-organisms can survive space travel in interplanetary conditions,
i.e., low temperature,
vacuum, UV radiation, and the shock experienced while impacting a solid planet
(Burchell et al 2001, Roten
et al 1998, O'Keefe et al 1996).
All these considerations are part of the hypothesis called panspermia
(Arrhenius 1903, Becquerel 1910a, 1910b) that there may be an
``interfertilisation'' between planets (Mileykowsky 1997, Portner et al 1961).

In the case of our own solar system at 2 AU, it is possible that
in the future (assuming the presence of liquid water
is encountered during a time scale compatible with the time for occurrence
of life on Earth) life may
appear on the planet Mars if some (presently hypothetical)
frozen liquid water can be melted.
Liquid water also could be produced on Europa as a consequence
of the evolution of the Sun. However, since the mass of Europa is relatively
small, it may be difficult for it to maintain an atmosphere
for a long time, and hence could evaporate into space.
A more massive body like Titan is perhaps better
adapted to host life (transit at 10 AU from several 10$^7$ years to about
10$^8$ years ).
This speculative scenario assumes that no change in the orbits of
the planets will occur
during the next 6$\times$10$^9$ years. From the point of view of dynamical
orbital changes related to stellar evolution, this assumption seems reasonable.

Effects of the stellar mass loss phenomena and tidal dissipation between the planets
and the red giant star are not negligible in the dynamics of the orbits
(Rasio et al. 1996). A Sun-like star will lose about 24$\%$ of its initial
mass during the RGB phase and about 20$\%$ during the AGB
phase (Schr\"oder and Sedlmayr 2001).
The transit of the habitable zone
that we have calculated, with a reasonably long duration for the
development of life, happens along the RGB, at a stage during which the
star has a radius less
than $\sim$ 100 R$_{\odot}$.
In Fig. 2 of Rasio et al. (1996), when a Sun like star reaches a radius
of about 100 R$_{\odot}$, only a little more than 10 $\%$ of the
initial mass of the star has been lost, and the effect of this mass loss
among the dynamical changes of the planet orbits (for planets more distant than 1 AU
not yet subject to tidal effects)
is to increase by 10 $\%$ the semi-major axis. The work of Rasio et al. (1996)
does consider the Reimers relation for representing the mass loss, according to
parameters used in Sackmann et al. (1993).
It is interesting to note that the increase of the semi-major axis, if occuring for
a planet during a time when it is in the habitable zone (itself expanding), will tend to
increase the time duration of habitable conditions for the planet.

\section{Possibility of detection by future interferometric projects}

\subsection{Available targets}

Planets around evolved stars of the class of sub-giant and red giant
may offer favorable conditions for hosting life at a few AU.
In the few next decades, the search for life around different type of stars
could be made using proposed space interferometers, such as the
Terrestrial Planet Finder (TPF) (Beichman, Woolf, \& Lindensmith 1999)
and Darwin (Fridlund \& Gondoin 2003).

Recently, a giant planet candidate around a subgiant class star has been discovered by
means of radial velocity measurements (Frink et al. 2002).
The detected planet has
M~$\sin~i$ = 8.9 M$_{Jup}$ (making it a planet candidate), and is orbiting at 1.3 AU from the
center of the star.
The parent star, HIP 75458,
is a K2III subgiant class star with log L/L$_{\odot}$ = 1.85
solar luminosity (Mallik 1999)
and with a mass of 1.05 M$_{\odot}$. It is located  31.5 pc from the Earth.
Figure 1 displays the habitable zone around HIP 75458.
The habitable zone, moving outward, is expanding to an outer limit of
14 AU and an inner limit at 7 AU. Habitable conditions
at 7 AU from the star have been encountered during
about 10$^8$ years, which is a reasonable, if optimistic, time for life to develop.
The presence of a Jupiter-like planet
encourages one to imagine that other bodies (perhaps even terrestrial ones)
could be orbiting this star.

Tables 2-4 display preliminary lists of evolved stars, and their characteristics,
contained within a sphere of radius 30 pc around our Sun. There
about 94 sources of luminosity class IV (Table 2), 44 sources of luminosity class
III (Table 3) and two sources of luminosity class II (Table 4). The luminosity class 
III and IV (respectively the giants and sub-giants classes) 
result from low and intermediate mass stars having just 
left the main sequence in their evolution. Luminosity class IV and III
stars are both of great interest.  In the first case, the habitable zone 
moves outward relatively slowly
and allows, for stars with masses $\leq$ 2 M$_{\odot}$, 
conditions that do not exclude the development of life.
In the second case, after the first ascent along the RGB and
the He flash, the stars evolve into a stable phase of core He burning
which is also favorable to the formation of life.  Approximately 
30 to 60 percent of the luminosity class IV stars in Table 3
are expected to be in this phase (Scalo \& Miller, 1979).
It is important to notice that luminosity class IV alone offers 94
nearby sources compared to the approximatly 1000 nearby main sequence stars
(of all spectral types).  

The ratio of all red giant stars (Tables 2-4) 
to the main sequence stars is thus about 14\%, which at first seems high considering the local space density of all giant stars 
is about $5 \times 10^{-3}$ compared to main sequence stars.  This 
value is based on main sequence and red giant star densities 
given in Cox (2000) and Neckel (1975).
However, a sample chosen based on a brightness or distance limit will select out
a larger fraction than expected on the basis of volume estimates because the
red giant stars are much brighter than the main sequence stars, and the
increase should be proportional to the ratio of the luminosities, or roughly
a factor of ten. 
Thus, within 30 pc, a reasonable sample has been shown to exist.
This sample allows us to consider evolved stars for the proposed future projects 
which may contribute to the understanding of the formation of life.

\subsection{Detection feasibility}

The development of large filled-aperture space telescopes coupled to efficient
coronagraphs and the development of space interferometers fed by large cooled telescopes, 
represent technological challenges that may be
overcome in the next two decades.  A first
step in the development of large space
telescopes like Darwin and TPF are precursor concepts for coronagraphy
(e.g. Eclipse, Trauger et al. 2003;  EPIC, Clampin et al. 2001) and
precursor concepts for interferometry like the FKSI mission concept 
(Fourier Kelvin Stellar Interferometer mission - Danchi et al 2003a, 2003b).
These instruments in the relatively near term could
directly observe some types of extrasolar planets around stars.

The difficulty of detecting a planet orbiting a red giant star
and characterizing its atmospheric signatures
lies in the very high contrast ratio
between the planet and the star. For a 1.0 M$_{\odot}$ star, due
to the squared dependence between the habitable zone size and the
stellar luminosity,
the habitable zone size associated with a star of luminosity
4 L$_{\odot}$ is of the order of 2 AU, while for a star of luminosity
25 L$_{\odot}$, the corresponding size is 5 AU, and, for
100 L$_{\odot}$, it is about 10 AU.
The contrast ratio required to observe a planet around a sub-giant or a
red giant star increases by a factor of 4 to 100
(range of luminosity corresponding
to a reasonably long duration of the transit) compared
to that required to observe a planet around main-sequence (FGK) dwarf stars 
(luminosity class V).
On the other hand the angular resolution needed to separate
the planet from the star is decreased due
the increased size of the habitable zone.  
For HIP 75458, the angular separation of the habitable zone
is about 0.3 arcsec from the star.
This separation is well within the resolution of filled-aperture space
telescopes with diameters in the 5-10 meter range operating
at visible and near-infrared wavelengths. 

We present a qualitative discussion below that shows 
how these factors approximately balance out in terms of the detectability
of planets around evolved stars.  A more detailed discussion is precluded 
at this time because the architectures for the TPF and Darwin missions 
are still under development, and a detailed detectability study requires 
detailed knowledge of instrumental parameters, which are not available at present.
We will present a more detailed treatment, based on simple model 
interferometers and coronagraphs in another paper. In the following 
discussion, for the purpose of developing
an appreciation of the detection issues, we 
use a main sequence star at a reference distance of 10 pc, and 
compare that to a red giant star at 30 pc.

Assuming the use of the FKSI concept or of any other future nulling interferometer,
three physical parameters representing the 
astrophysical source are of importance
in order to estimate the potential of detection. These physical parameters are:
(1) the angular distance between the planet and the star that has to be
compared to the inner working angle (IWA) of the instrument;
(2) the contrast ratio between
the planet and the star;
(3) and the stellar diameter, which affects the stellar leakage from a perfect
optimum nulling. 
For a 1 M$_{\odot}$ star, with the 
habitable zone varying between 
2 and 15 AU (middle in the width), 
the corresponding stellar radius variations will be 
approximately between 2 and 15 R$_{\odot}$ (see Figure 1(a) and 1(b);
note this is an approximation because the effective temperature of the star varies also). 

We now discuss the relative difficulty of detecting a planet around red giant
stars compared to main sequence stars.  The reason for this comparison is that 
the designs for TPF and Darwin will be based on the requirement to detect earth-like
planets around the habitable zone of nearby main sequence (F,G,K) dwarf stars (luminosity
class V).  Thus we show that if TPF and Darwin can detect earth-like planets
around these stars, then these instruments will also be able to detect such planets
around subgiant and red giant stars, i.e., luminosity classes IV and III.

The inner working angle (IWA) of a coronagraph is:

\begin{equation}
\label{ }
{\rm IWA_C} \sim 4 ( \lambda / D )
\end{equation}

where $\lambda$ and D are the wavelength of light and the diameter of the
primary mirror of the telescope, respectively.
The corresponding IWA for a simple nulling interferometer is:

\begin{equation}
\label{ }
{\rm IWA_I} \sim ( \lambda / 2B )
\end{equation}

where B is the baseline length and $\lambda$ was defined previously.

For the coronagraph systems, $\rm IWA_C$ varies from approximately 0.32 arcsec for
a 1.8 m primary mirror and a wavelength of 0.7 $\mu$m 
as planned for the EPIC and Eclipse missions, and is 0.096 arcsec
for the 6 m longest dimension for the planned mirror for the coronagraphic
version of TPF, now called TPF-C.  Despite the longer wavelength of light
for the infrared interferometers, $\rm IWA_I$ is typically much smaller than
$\rm IWA_C$.  For the FKSI system, $\rm IWA_I$ is 0.043 arcsec at a center wavelength
of 5 $\mu$m, while $\rm IWA_I$ is 0.029 arcsec for a 35 m baseline on a 
structurally connected interferometer, or 0.017 arcsec for a 60 m baseline
on a free flying interferometer, both assuming a center wavelength of 
10 $\mu$m.

These IWAs can now be compared to the angular size of the habitable zones
for the two extreme cases for the red giant stars, namely 2 AU and 15 AU.
At a reference distance of 10 pc the habitable zone is located at 0.20 arcsec from the star and 1.5 
arcsec for the 2 AU and 15 AU cases, respectively.  For a star at a reference
distance of 30 pc, the angular sizes are proportionally smaller, 0.067 arcsec and 0.50
arcsec, respectively.

Based on the IWAs calculated here, the IWA of the small pathfinding 
coronagraphs like Eclipse and EPIC will be adequate only for the nearest
such stars, while the larger TPF-C has resolution sufficient to resolve
the habitable zone for 
essentially all the red giant stars.  The interferometer IWAs are 
all smaller than the angular size of the habitable zone for all the cases
discussed here, which means that this instrumental parameter is 
favorable for the detection of planets around red giant stars.

The planet-star intensity contrast ratio in reflected light is 
approximately:

\begin{equation}
\label{ }
I_p / I_s \propto {( R_p / R_{ps} )} ^ 2
\end{equation}

where $I_p$ and $I_s$ are the planet and star intensities, respectively,
while $R_p$ is the radius of the planet and $R_{ps}$ is the distance
from the star to the planet (cf. Charbonneau et al. 1999).  

For infrared radiation emitted by the planet in thermal equilibrium with the
star, the
planet-star intensity ratio is:

\begin{equation}
\label{ }
I_p / I_s \propto {( R_p / R_{s} )} ^ 2 \left[ { \exp { ( h \nu / kT_s )} - 1 } 
\over  {\exp { (h \nu / kT_p )} - 1 } \right]
\end{equation}

where $T_s$ is the effective temperature of the star, and $T_p$ is the
equilibrium temperature of the planet, as given by Eqn. (1) previously.
The quantities $R_p$ and $R_s$ are the radii of the planet and star,
respectively.

In reflected light, the planet-star contrast ratio is reduced in comparison
with main sequence stars because of the much larger planet-star separation, which 
amounts to factors of 4 to 225 for the 2 AU and 15 AU cases, respectively.
In the thermal infrared, the contrast ratios are reduced 
by factors of approximately 4 to 100 (compared to those of main
sequence stars) as the radius of the star varies
from about 2$R_{\odot}$ to 10$R_{\odot}$ over the course of the evolution
of the star during the red giant phase.

An additional consideration for the nulling interferometer is the stellar
leakage mentioned previously.  For a simple Bracewell interferometer
with a $\theta ^2$ null, the stellar leakage is given by:

\begin{equation}
\label{ }
L_s \propto  {\bigg({ {\theta}_s \over  {\lambda / B } }\bigg) ^2 }
\end{equation}

Thus the stellar leakage is proportional to the square of the angular
size of the star divided by the angular resolution of the interferometer,
or equivalently, the fringe spacing.
This means the stellar leakage for a giant star of radius 2$R_{\odot}$ at 20 pc is
the same as that of a 1$R_{\odot}$ main sequence star at 10 pc.  However,
the observable red giant stars are on average further away from our solar system
than the main sequence stars, and the increase in the stellar leakage signal 
is expected to be at the low end of the range, or comparable to that of
the main sequence stars.  

We can summarize our considerations of the detectability of these
stars by noting that the problem is only moderately more difficult 
by a factor of 4 to 10 in terms of the star-planet contrast ratio, which
is in the range of the variation expected for the most optimistic to 
the worst cases for the main sequence stars, given the variations in spectral type
and planet size being considered for TPF/Darwin.  The inner working
angles of the coronagraphs and interferometers are adequate for most of the 
stars in the tables in this paper.  The stellar leakage is approximately
equal or only modestly worse than that for main sequence stars since the giant stars
are on average at larger distances from our solar system.
More detailed studies of detectability issues are planned as more 
becomes known about the architectures for TPF and Darwin.

\section{Discussion and Conclusions}

We have estimated the time transit of the habitable zone
around evolved stars of initial mass 1.0, 1.5 and 2.0 M$_{\odot}$ from
the time period after which a star moves off the main sequence until it
reaches the point of the He flash.  After the He flash, there 
is an additional potential phase for life  during the period
of quiescent core He burning prior to the star's ascent along
the Asymptotic Giant Branch.
The case of a 0.85 M$_{\odot}$ star was not analyzed because
the time for such a star to leave the main sequence is
longer than the age of the Universe.
A simple hypothesis was used to estimate the inner and
outer limits of the habitable zone, i.e., a  blackbody assumption and
extrapolation of results presented in Kasting et al. (1993) defining
the habitable zone limits around a solar type star.
In our approximation the fact that the spectrum of the evolved star differs
from the solar spectrum used in Kasting et al. (1993) is neglected.
Only the luminosity of the stars is considered,
while the planet is assumed to be a greybody with perfect absorption at any
wavelength.

We conclude that life may exist around evolved stars
of the sub-giant and red giant classes at several AU from the stars.
Although the temporal evolution of a star is rapid after leaving the main sequence,
the temporal transit of the habitable zone does not appear incompatible
with the possible duration for the development of life.
Moreover, micro-organisms could be transported from
a  planet where life is ending to a planet where favorable illumination
conditions for the rebirth of life are encountered. This type of 
mechanism may be unique, or may be encountered exclusively
in the environment of evolved stars.

We speculate also that around our Sun, life
may develop in the future at distances in the range 2-9 AU.
This is deduced from the the temporal transit of the habitable zone, which
is estimated to be of the order of several 10$^9$ years for a planet at 2 AU
and around 10$^8$ years for a planet at 9 AU.
Furthermore, after the Sun ascends the Red Giant Branch for the
first time and undergoes the He flash, there is an additional 
long stable period of core He burning $\sim$ $10^9$ years during
which life can evolve in the habitable zone, which now is 
from 7 to 22 AU from the star.

Nearly 150 evolved stars  (sub-giants and red giants)
are within 30 pc of our solar system (see the details in Tables 2-4).
At the present time we do not know
if life is common in our Universe, or extremely rare, or even unique on Earth.
If future
observing instruments allow the detection of signatures of
life on various planets, then
the observation of planets, possibly hosting life, around evolved stars
may help to set some limit on the time required for life to develop.
For example, if life is found on a planet inside a habitable zone
transiting in 10$^7$ years, this proves that life could develop within
this period, setting a much smaller upper limit to the formation
of life than presently known.
In addition, if
life appears to be common, another
interesting case would be to find a planet containing
liquid water within an habitable zone of rather fast transition
(like 10$^7$ years) with no
signature of life. This case (fast conditions of habitability) could
contribute to set new lower limits for the emergence of life.
The environments of evolved stars may represent some interesting test cases.

These test cases can consider the fundamental hypothesis that early life may have been connected to volcanic environments such as
deep sea hydrothermal vents. Thermophilic life may populate the ocean of superficially frozen planets. The expansion of
the habitable zone around the star in evolution will change the habitable conditions  
of the surface of the planet by increasing
its temperature, by increasing the rate of water evaporation and 
thus by triggering the hydrological cycles (evaporation, clouds
formation, rain process). If life was already present underneath the oceans, 
it could spread on the ground layer very fast.
An interesting issue to tackle by observing a freshly habitable planet just 
entering a 'rapidly' expanding habitable zone is the coupling
between the development of life in the ocean and its development on the ground layer. 
Life is known to have developed first in
the Earth's oceans. On other planets, if life as we know it is found, one may wonder: 
is the chlorophyl signature (mostly representing the ground layer activity) always 
encountered when bioatmospheric signatures (related to the global activity: oceans + ground layer)
will be found? A rapidly expanding habitable zone could help improve our understanding of the 
time delay coupling between these two different bio-environments: ocean and surface layer.

We conclude that future
observing programs focused on the search for planets with possible
signatures of life should not omit the
class of sub-giants and red giant stars. The study of
objects of this class will further contribute to
the understanding of the formation of life in the Universe.
Filled-aperture space telescopes with
diameters in the range of 5 to 10 meters operating at
visible and near-infrared wavelengths and space-based mid-infrared interferometers
with baselines in the range from 35 to 60 m, provide adequate
resolution and sensitivity to detect earth-like planets around 
red giant stars.

\acknowledgments

We thank P. Baudoz, F. Selsis, T. Guillot and F. Th\'evenin for useful
discussions and K.-P. Schr\"oder for constructive remarks on an earlier
version of this manuscript.  We thank Dr. J. Rajagopal for a 
careful reading of the manuscript.  We also appreciate constructive comments on
the manuscript from our referee, Dr. John Scalo.

\clearpage
\begin{figure}
\plotone{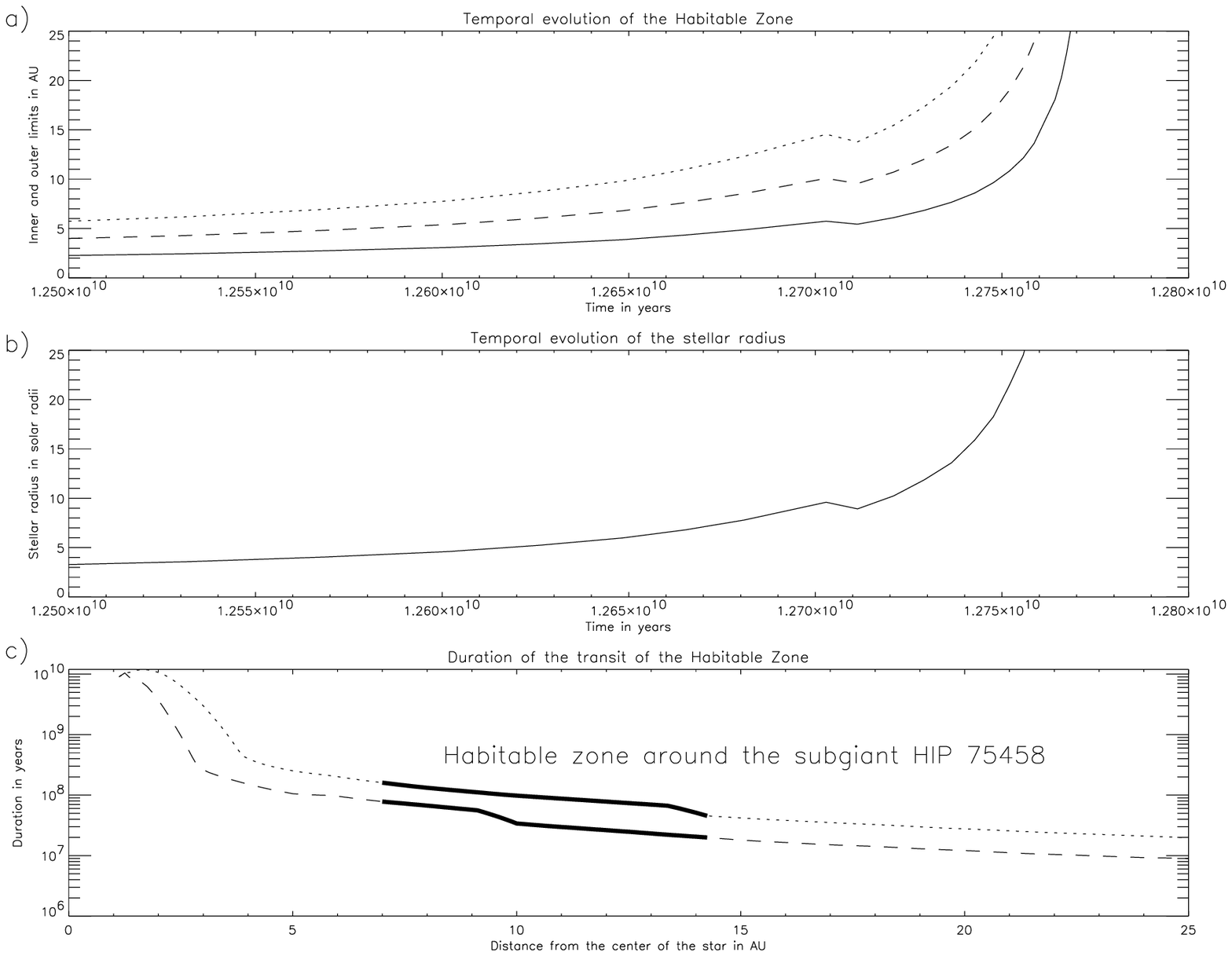} 
\caption{\label{sketch} a) The evolution of inner
and outer limit of the habitable zone around a post-main sequence
star of 1.0 M$_{\odot}$. The time scale displayed is between 12.5
and 12.8 Gyrs. b)~The corresponding evolution of the stellar radius 
(from Maeder and Meynet 1988).
c) The duration of the transit of the habitable
zone represented at different distances from the star.}
\end{figure}

\clearpage
\begin{figure}
\plotone{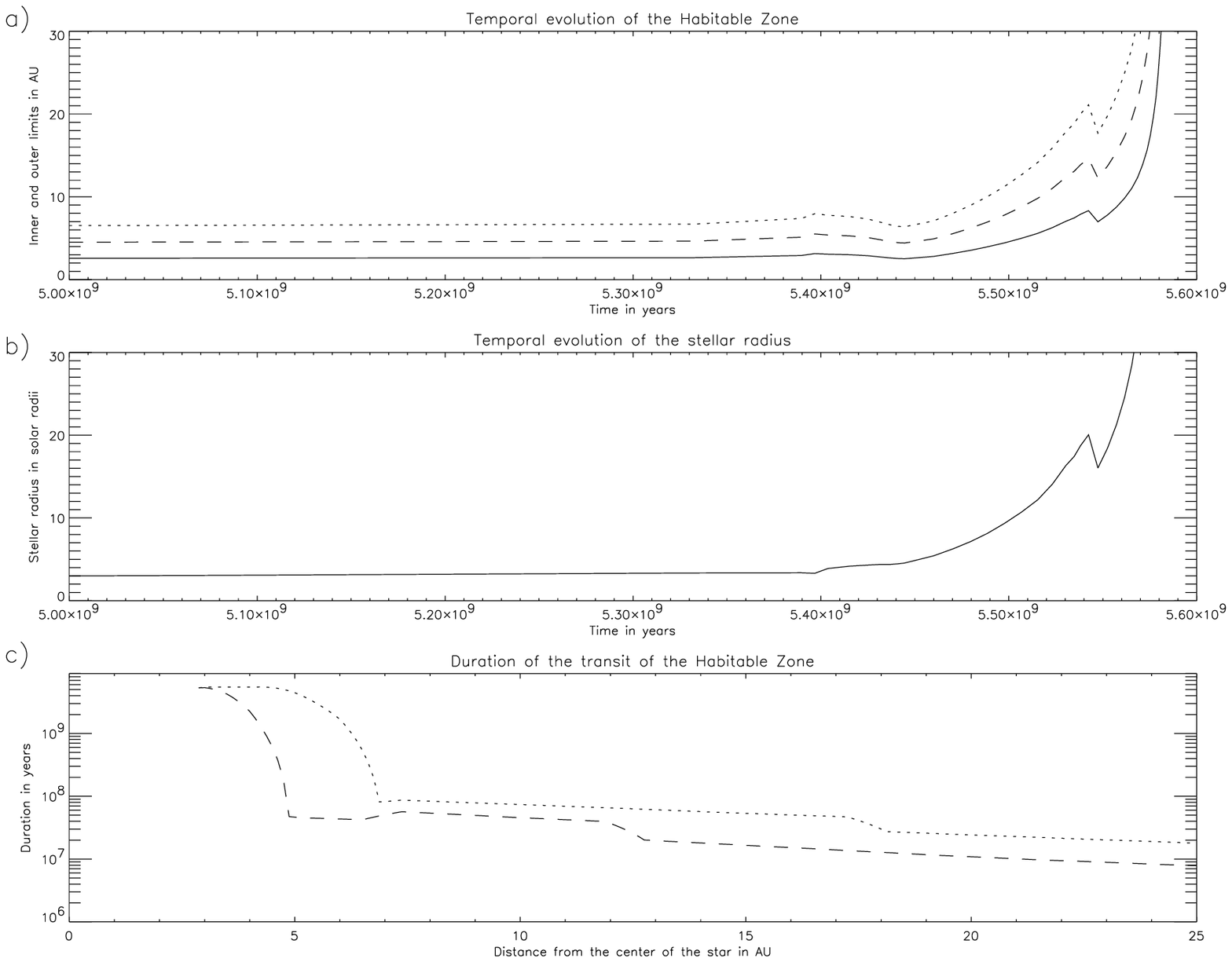} 
\caption{\label{sketch} a) The evolution of inner
and outer limit of the habitable zone around a star of 1.5
M$_{\odot}$. The time scale displayed is between 5.0 and 5.6 Gyrs.
b) The corresponding evolution of the stellar radius 
(from Maeder and Meynet 1988).
c) The duration of the transit of the habitable zone represented
at different distances from the star.}
\end{figure}

\clearpage
\begin{figure}
\plotone{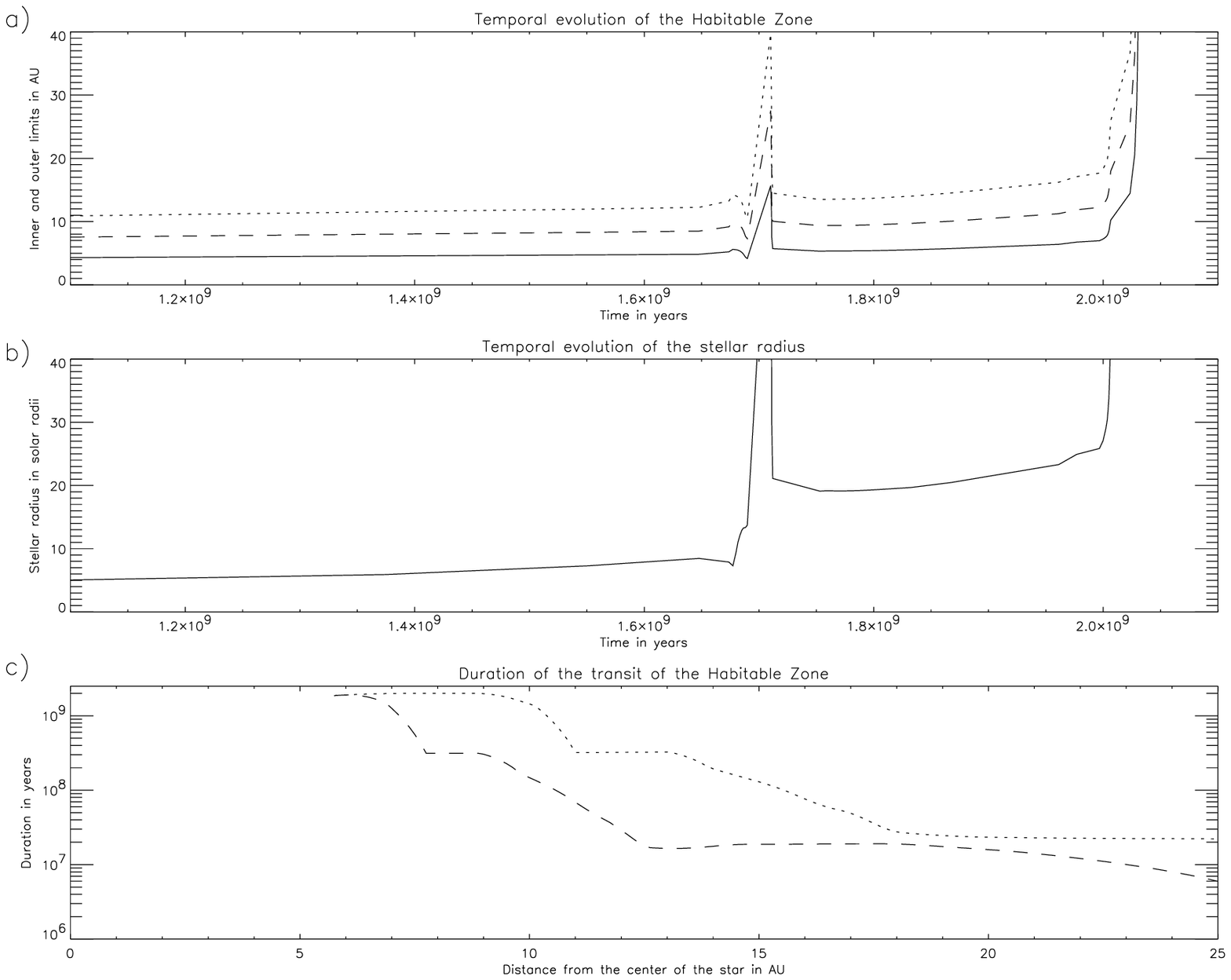} 
\caption{\label{sketch} a) The evolution of inner
and outer limit of the habitable zone around a star of 2.0
M$_{\odot}$. The time scale displayed is between 1.5 and 2.1 Gyrs.
b) The corresponding evolution of the stellar radius 
(from Maeder and Meynet 1988).
c) The duration of the transit of the habitable zone represented
at different distances from the star.}
\end{figure}

\clearpage

\begin{deluxetable}{ccccc}
\tabletypesize{\scriptsize}
\tablecolumns{5}
\tablewidth{0pc}
\tablecaption{Lifetimes (Gyr) of stages of stellar evolution as a function of initial stellar mass}
\tablehead{ \colhead{Mass (M$_{\odot}$}    &
\colhead{Main Sequence} &   \colhead{Subgiant}     
& \colhead{First Red Giant} & \colhead{Core He Burning}   }
\startdata

1.0 & 7.41 & 2.63 & 1.45  & 0.95  \\
1.5 & 1.72 & 0.41 & 0.18  & 0.26  \\
2.0 & 0.67 & 0.11 & 0.04  & 0.10  \\
\enddata
\end{deluxetable}

\clearpage

\begin{deluxetable}{llrrrrrcr}
\tabletypesize{\scriptsize}
\tablecolumns{9}
\tablewidth{0pc}
\tablecaption{Luminosity Class IV Stars Within 30 pc}
\tablehead{ \colhead{Name}    &
\colhead{Alt. Name} &   \colhead{Type\tablenotemark{a}}     & \colhead{RA
J(2000)} & \colhead{Dec J(2000)}    & \colhead{B\tablenotemark{b}}   & \colhead{V\tablenotemark{c}}
& \colhead{Spec. Type} & \colhead{Parallax\tablenotemark{d}} }
\startdata
HD 142                &           &    & 00 06 19.18  & -49 04 30.7    &  6.22    &   5.70 & G1   &  39.0   \\
$\alpha$ And          & HD 358    & SB & 00 08 23.26  & +29 05 25.6    &  1.95    &   2.06 & B8   &  33.6   \\
$\beta$ Cas           & HD 432    & dS & 00 09 10.69  & +59 08 59.2    &  2.61    &   2.27 & F2   &  59.9   \\
HD 2071               &           &    & 00 24 42.55  & -53 59 02.4    &  8.0     &   7.4  & G8   &  36.2   \\
$\beta$ Hyi           & HD2151    &  V & 00 25 45.07  & -77 15 15.3    &  3.42    &   2.80 & G2   & 133.9   \\
HD 3141               &           &    & 00 35 00.87  & +42 41 41.2    &  8.87    &   8.01 & K0   &  34.5   \\
$\kappa$ Tuc          & HD 7788   &  V & 01 15 46.16  & -68 52 33.3    &  4.73    &   4.25 & F6   &  48.9   \\
$\omega$ And          & HD 8799   &    & 01 27 39.38  & +45 24 24.1    &  5.25    &   4.83 & F5   &  35.3   \\
HD 9562               &           &    & 01 33 42.84  & -07 01 31.2    &  6.37    &   5.76 & G2   &  33.7   \\
HD 10086              &           &    & 01 39 36.02  & +45 52 40.0    &  7.30    &   6.61 & G5   &  46.7   \\
$\epsilon$ Scl        & HD 10830  & Ds & 01 45 38.76  & -25 03 09.4    &  5.70    &   5.31 & F2   &  36.5   \\
$\alpha$   Tri        & HD 11443  & SB & 01 53 04.91  & +29 34 43.8    &  3.90    &   3.41 & F6   &  50.9   \\
HD 13825              &           &    & 02 15 24.41  & +24 16 16.7    &  7.50    &   6.81 & G8   &  37.9   \\
AG+01 242             & HD 14214  & SB & 02 18 01.44  & +01 45 28.1    &  6.18    &   5.58 & G0.5 &  40.0   \\
HD 16417              &           &    & 02 36 58.61  & -34 34 40.7    &  6.45    &   5.78 & G5   &  39.2   \\
NSV 893               & HD 16765  &  V & 02 41 14.00  & -00 41 44.4    &  6.23    &   5.71 & F7   &  46.2   \\
87 Cet                & HD 17094  & dS & 02 44 56.54  & +10 06 50.9    &  4.58    &   4.20 & F0   &  38.7   \\
HD 17190              &           &    & 02 46 15.21  & +25 38 59.6    &  8.66    &   7.81 & K1   &  39.0   \\
$\alpha$ For          & HD 20010  &    & 03 12 04.53  & -28 59 15.4    &  4.36    &   3.85 & F8   &  70.9   \\
$\delta$ Eri          & HD 23249  & RS & 03 43 14.90  & -09 45 48.2    &  4.43    &   3.51 & K0   & 110.6   \\
V891 Tau              & HD 26913  & BY & 04 15 25.79  & +06 11 58.7    &  7.60    &   6.96 & G5   &  47.9   \\
V774 Tau              & HD 26923  &  V & 04 15 28.80  & +06 11 12.7    &  6.87    &   6.33 & G0   &  47.2   \\
13 Ori                & HD 33021  &    & 05 07 38.31  & +09 28 18.4    &  6.79    &   6.17 & G1   &  35.3   \\
HD 37986              &           &    & 05 41 53.69  & -15 37 50.0    &  8.16    &   7.37 & G8/K0&  36.1   \\
HD 39881              &           &    & 05 56 03.43  & +13 55 29.7    &  7.25    &   6.60 & G5   &  35.7   \\
$\beta$ Aur           &           & Al & 05 59 31.72  & +44 56 50.8    &  1.97    &   1.89 & A2   &  39.7   \\
HD 45184              &           &    & 06 24 43.88  & -28 46 48.4    &  7.01    &   6.37 & G2   &  45.4   \\
31 Gem                & HD 43737  &  V & 06 45 17.36  & +12 53 44.1    &  3.79    &   3.40 & F5   &  57.0   \\
HD 50806              &           &    & 06 53 33.94  & -28 32 23.3    &  6.75    &   6.04 & G5   &  38.8   \\
QW Pup                & HD 55892  & a2 & 07 12 33.63  & -46 45 33.5    &  4.80    &   4.50 & F0   &  47.2   \\
55 Gem A              & HD 56986  & SB & 07 20 07.38  & +21 58 56.4    &  3.87    &   3.53 & F0   &  55.5   \\
HD 62644              &           &    & 07 42 57.10  & -45 10 23.2    &  5.84    &   5.04 & G6   &  41.4   \\
HD 63433              &           &    & 07 49 55.06  & +27 21 47.5    &  7.57    &   6.93 & G5   &  45.8   \\
10 Cnc                & HD 67228  &    & 08 07 45.86  & +21 34 54.5    &  5.91    &   5.30 & G1   &  42.9   \\
13 UMa                & HD 78154  & Ds & 09 10 23.55  & +67 08 02.5    &  5.29    &   4.80 & F6   &  48.9   \\
$\psi$ Vel            & HD 82434  &    & 09 30 42.00  & -40 28 00.4    &  3.91    &   3.59 & F3   &  53.9   \\
23 UMa                & HD 81937  &  V & 09 31 31.71  & +63 03 42.7    &  3.97    &   3.65 & F0   &  43.2   \\
25 UMa                & HD 82328  & SB & 09 32 51.43  & +51 40 38.3    &  3.63    &   3.20 & F6   &  74.2   \\
40 Leo                & HD 89449  & dS & 10 19 44.17  & +19 28 15.3    &  5.24    &   4.80 & F6   &  47.2   \\
I Car                 & HD 90589  &  V & 10 24 23.71  & -74 01 53.8    &  4.32    &   3.99 & F2   &  61.7   \\
p Vel                 & HD 92139  & SB & 10 37 18.14  & -48 13 32.2    &  4.14    &   3.84 & F4   &  37.7   \\
78 Leo                & HD 99028  & SB & 11 23 55.45  & +10 31 46.2    &  4.35    &   4.00 & F4   &  41.3   \\
83 Leo                & HD 99491  &    & 11 26 45.32  & +03 00 47.2    &  7.29    &   6.49 & K0   &  56.6   \\
HD 104304             &           &    & 12 00 44.45  & -10 26 45.6    &  6.32    &   5.54 & G9   &  77.5   \\
HD 114174             &           &    & 13 08 51.02  & +05 12 26.1    &  7.47    &   6.80 & G5   &  38.1   \\
HD 114837             &           &    & 13 14 15.14  & -59 06 11.6    &  5.40    &   4.92 & F7   &  55.5   \\
HD 118096             &           &    & 13 34 03.24  & +33 13 42.2    & 10.35    &   9.23 & K5   &  43.7   \\
1 Cen                 & HD 224481 & SB & 13 45 41.25  & -33 02 37.4    &  4.61    &   4.23 & F3   &   6.7   \\
$\tau$ Boo            & HD 120136 &  V & 13 47 15.74  & +17 27 24.9    &  4.98    &   4.50 & F6   &  64.1   \\
$\eta$ Boo            & HD 121370 & SB & 13 54 41.08  & +18 23 51.8    &  3.26    &   2.68 & G0   &  88.2   \\
HD 122862             &           &    & 14 08 27.17  & -74 51 01.0    &  6.60    &   6.02 & G2.5 &  34.9   \\
$\iota$ Vir           & HD 124850 &  V & 14 16 00.87  & -06 00 02.0    &  4.60    &   4.10 & F7   &  46.7   \\
18 Boo                & HD 125451 & Ds & 14 19 16.28  & +13 00 15.5    &  5.764   &   5.40 & F5   &  38.3   \\
9 Lib                 & HD 130841 &  V & 14 50 52.71  & -16 02 30.4    &  2.912   &   2.75 & A3   &  42.3   \\
HD 136064             &           &    & 15 14 38.34  & +67 20 48.2    &  5.66    &   5.10 & F9   &  39.5   \\
HD 134606             &           &    & 15 15 15.04  & -70 31 10.6    &  7.60    &   6.86 & G5   &  37.2   \\
37 Lib                & HD 138716 &    & 15 34 10.70  & -10 03 52.3    &  5.63    &   4.61 & K1   &  34.5   \\
NSV 7350              & HD 142860 &  V & 15 56 27.18  & +15 39 41.8    &  4.33    &   3.85 & F6   &  89.9   \\
13 Dra                & HD 144284 &    & 16 01 53.35  & +58 33 54.9    &  4.53    &   4.01 & F8   &  47.8   \\
$\xi$ Sco A           & HD 144070 & Ds & 16 04 22.3   & -11 22 20      &  5.24    &   4.77 & F5   &    36   \\
NSV 7915              & HD 150680 & SB & 16 41 17.16  & +31 36 09.8    &  3.54    &   2.89 & G0   &  92.6   \\
HD 152311             &           &    & 16 53 25.22  & -20 24 56.0    &  6.53    &   5.87 & G5   &  35.7   \\
$\delta$ Her          & HD 156164 & Ds & 17 15 01.91  & +24 50 21.1    &  3.22    &   3.12 & A3   &  41.6   \\
HD 157347             &           &    & 17 22 51.29  & -02 23 17.4    &  6.94    &   6.29 & G5   &  51.4   \\
HD 158332             &           &    & 17 27 34.58  & +26 47 41.9    &  8.54    &   7.73 & K1   &  33.4   \\
$\lambda$ Ara         & HD 160032 &    & 17 40 23.83  & -49 24 56.1    &  5.17    &   4.77 & F3   &  45.7   \\
86 Her                & HD 161797 &    & 17 46 27.53  & +27 43 14.4    &  4.17    &   3.41 & G5   & 119.1   \\
$\zeta$ Ser           & HD 150680 &    & 18 00 29.01  & -03 41 25.0    &  4.99    &   4.63 & F2   &  92.6   \\
72 Oph                & HD 165777 & Ds & 18 07 20.98  & +09 33 49.8    &  3.86    &   3.72 & A4s  &  39.4   \\
SAO 85892             & HD 168874 & Ds & 18 20 49.26  & +27 31 49.2    &  7.59    &   7.00 & G2   &  34.0   \\
HD 177565             &           &    & 19 06 52.46  & -37 48 38.4    &  6.88    &   6.16 & G5   &  58.2   \\
31 Aql                & HD 182572 &  V & 19 24 58.20  & +11 56 39.9    &  5.93    &   5.16 & G8   &  66.0   \\
$\delta$ Aql          & HD 182640 &  V & 19 25 29.90  & +03 06 53.2    &  3.68    &   3.40 & F0   &  65.1   \\
HD 183877             &           &    & 19 32 40.33  & -28 01 11.3    &  7.83    &   7.15 & G5   &  38.4   \\
$\beta$ Aql           & HD 188512 &  V & 19 55 18.79  & +06 24 24.3    &  4.57    &   3.71 & G8   &  73.0   \\
HD 190360             &           &    & 20 03 37.41  & +29 53 48.5    &  6.44    &   5.71 & G6   &  62.9   \\
HD 190771             &           &    & 20 05 09.78  & +38 28 42.4    &  6.82    &   6.17 & G5   &  54.0   \\
27 Cyg                & HD 191026 & RS & 20 06 21.77  & +35 58 20.9    &  6.21    &   5.36 & K0   &  41.3   \\
$\delta$ Pav          & HD 190248 &  V & 20 08 43.61  & -66 10 55.4    &  4.32    &   3.56 & G7   & 163.7   \\
HD 195564             &           &    & 20 32 23.70  & -09 51 12.2    &  6.34    &   5.65 & G2.5 &  41.3   \\
6 Del                 & HD 196524 & SB & 20 37 32.94  & +14 35 42.3    &  4.04    &   3.63 & F5   &  33.5   \\
$\eta$ Cep            & HD 198149 &    & 20 45 17.38  & +61 50 19.6    &  4.35    &   3.41 & K0   &  69.7   \\
$\tau$ Cyg            & HD 202444 &  V & 21 14 47.49  & +38 02 43.1    &  4.11    &   3.72 & F0   &  47.8   \\
$\alpha$ Cep          & HD 203280 &  V & 21 18 34.77  & +62 35 08.1    &  2.66    &   2.44 & A7   &  66.8   \\
HD 203985             &           &    & 21 27 01.33  & -44 48 30.8    &  8.38    &   7.48 & K1   &  42.5   \\
51 Cap                & HD 207958 &    & 21 53 17.77  & -13 33 06.4    &  5.45    &   5.08 & F3   &  36.2   \\
$\epsilon$ Cep        & HD 211336 & dS & 22 15 02.19  & +57 02 36.9    &  4.47    &   4.19 & F0   &  38.9   \\
HD 212330             &           &    & 22 24 56.39  & -57 47 50.7    &  5.99    &   5.31 & G3   &  48.8   \\
49 Peg                & HD 216385 &    & 22 52 24.08  & +09 50 08.4    &  5.64    &   5.16 & F7   &  37.3   \\
$\rho$ Ind            & HD 216437 &    & 22 54 39.48  & -70 04 25.4    &  6.69    &   6.06 & G2.5 &  37.7   \\
HD 217107             &           &    & 22 58 15.54  & -02 23 43.4    &  6.90    &   6.18 & G8   &  50.7   \\
6 And                 & HD 218804 &    & 23 10 27.20  & +43 32 39.2    &  6.38    &   6.00 & F5   &  35.4   \\
35 Cep                & HD 222404 &  V & 23 39 20.85  & +77 37 56.2    &  4.26    &   3.22 & K1   &  72.5   \\
HD 224022             &           &  P & 23 54 38.62  & -40 18 00.2    &  6.60    &   6.03 & F8   &  35.9   \\
\enddata
\tablenotetext{a}{Type of system.  a2 = $\alpha$2 CVn; Al = Algol; BY = BY Dragonis; dS = delta Scuti;
Ds = double or multiple star; Pu = pulsating variable; Ro = rotating variable; RS = RS CVn; V = variable }
\tablenotetext{b}{magnitudes}
\tablenotetext{c}{magnitudes}
\tablenotetext{d}{milliarcsecs (mas) }
\end{deluxetable}


\begin{deluxetable}{llrrrrrcr}
\tabletypesize{\scriptsize}
\tablecolumns{9}
\tablewidth{0pc}
\tablecaption{Luminosity Class III Stars Within 30 pc}
\tablehead{ \colhead{Name}    &
\colhead{Alt. Name} &   \colhead{Type}     & \colhead{RA
J(2000)} & \colhead{Dec J(2000)}    & \colhead{B}   & \colhead{V}
& \colhead{Spec. Type}  & \colhead{Parallax} }
\startdata

$\alpha$ Phe          & HD 2261   & SB & 00 26 17.05  & -42 18 21.5    &  3.46    &   2.37 & K0   &  42.1   \\
$\beta$ Cet           & HD 4128   &  V & 00 43 35.37  & -17 59 11.8    &  3.06    &   2.04 & K0   &  34.0   \\
$\chi$ Cet            & HD 11171  & Ds & 01 49 35.10  & -10 41 11.1    &  4.981   &   4.66 & F3   &  42.4   \\
$\alpha$ Ari          & HD 12929  &  V & 02 07 10.41  & +23 27 44.7    &  3.15    &   2.00 & K2   &  49.5   \\
$\iota$ Hyi           & HD 21024  &    & 03 15 57.66  & -77 23 18.4    &  5.914   &   5.51 & F4   &  34.0   \\
$\gamma$ Dor          & HD 27290  & Pu & 04 16 01.59  & -51 29 11.9    &  4.55    &   4.20 & F4   &  49.3   \\
$\alpha$ Tau          & HD 29139  &  V & 04 35 55.24  & +16 30 33.5    &  2.39    &   0.85 & K5   &  50.1   \\
HD 29883              &           &    & 04 43 35.44  & +27 41 14.6    &  8.92    &   8.01 & K5   &  44.7   \\
67 Eri                & HD 33111  &  V & 05 07 50.99  & -05 05 11.2    &  2.92    &   2.79 & A3   &  36.7   \\
HD 290054             &           &    & 05 14 48.13  & +00 39 43.1    & 11.1     &   9.9  & K2   &  38.1   \\
$\alpha$ Aur          & HD 34029  & RS & 05 16 41.36  & +45 59 52.8    &  0.88    &   0.08 & G5   &  77.3   \\
CAL 42                & HD 36705  & Ro & 05 28 44.83  & -65 26 54.9    &  7.73    &   6.93 & K1   &  66.9   \\
$\beta$ Col           & HD 39425  &    & 05 50 57.59  & -35 46 05.9    &  4.28    &   3.12 & K2   &  37.9   \\
SV* ZI 543            & HD 47205  &  V & 06 36 41.04  & -19 15 21.2    &  5.01    &   3.96 & K1   &  50.4   \\
$\alpha$ Cha          & HD 71243  &    & 08 18 31.55  & -76 55 11.0    &  4.436   &   4.06 & F5   &  51.4   \\
q Pup                 & HD 63744  &    & 08 18 33.31  & -36 39 33.4    &  4.657   &   4.44 & A7   &  14.4   \\
SV LMi                & HD 82885  & RS & 09 35 39.50  & +35 48 36.5    &  6.18    &   5.41 & G8   &  89.5   \\
46 LMi                & HD 94264  &  V & 10 53 18.71  & +34 12 53.5    &  4.87    &   3.83 & K0   &  33.4   \\
U Car                 & HD 94510  &  V & 10 53 29.66  & -58 51 11.4    &  4.713   &   3.79 & K1   &  33.7   \\
$\eta$ Cru            & HD 105211 &  Ds& 12 06 52.90  & -64 36 49.4    &  4.462   &   4.14 & F2   &  50.8   \\
$\gamma$ Cru          & HD 108903 &  V & 12 31 09.96  & -57 06 47.6    &  3.22    &   1.63 & M3.5 &  37.1   \\
$\delta$ Mus          & HD 112985 & SB & 13 02 16.26  & -71 32 55.9    &  4.80    &   3.62 & K2   &  35.9   \\
GJ 9427               &SAO 63355  &    & 13 07 35.06  & +34 24 06.1    & 10.53    &   9.30 & K3   &  41.9   \\
HD 118036             &           &    & 13 34 16.26  & -00 18 49.6    &  8.27    &   7.38 & K4   &  41.0   \\
$\alpha$ Boo          & HD 124897 &  V & 14 15 39.67  & +19 10 56.7    &  1.19    &  -0.04 & K1.5 &  88.9   \\
$\gamma$ Boo          & HD 127762 & dS & 14 32 04.67  & +38 18 29.7    &  3.23    &   3.00 & A7   &  38.3   \\
107 Vir               & HD 129502 &    & 14 43 03.62  & -05 39 29.5    &  4.26    &   3.90 & F2   &  53.5   \\
HD 130322             &           &    & 14 47 32.73  & -00 16 53.3    &  8.80    &   8.05 & K0   &  33.6   \\
$\beta$ TrA           & HD 141891 &    & 15 55 08.56  & -63 25 50.6    &  3.14    &   2.85 & F2   &  81.2   \\
26 Sco                & HD 151680 &  V & 16 50 09.81  & -34 17 35.6    &  3.44    &   2.29 & K2.5 &  49.9   \\
$\kappa$ Oph          & HD 153210 &  V & 16 57 40.10  & +09 22 30.1    &  4.35    &   3.20 & K2   &  38.0   \\
$\alpha$ Oph          & HD 159561 &  V & 17 34 56.07  & +12 33 36.1    &  2.23    &   2.10 & A5   &  69.8   \\
60 Oph                & HD 161096 &  V & 17 43 28.35  & +04 34 02.3    &  3.967   &   2.77 & K2   &  39.8   \\
10 Sgr                & HD 165135 &  V & 18 05 48.49  & -30 25 26.7    &  3.99    &   2.99 & K0   &  33.9   \\
111 Her               & HD 173880 & Ds & 18 47 01.27  & +18 10 53.5    &  4.477   &   4.34 & A5   &  35.2   \\
38 Sgr                & HD 176687 & Ds & 19 02 36.71  & -29 52 48.4    &  2.704   &   2.60 & A2   &  36.6   \\
HD 187237             &           &    & 19 48 00.88  & +27 52 10.3    &  7.51    &   6.88 & G2   &  38.4   \\
HD 195627             &           &    & 20 35 34.85  & -60 34 54.3    &  5.031   &   4.75 & F1   &  36.3   \\
$\varepsilon$ Cyg       & HD 197989 &    & 20 46 12.68  & +33 58 12.9    &  3.49    &   2.50 & K0   &  45.3   \\
$\nu$ Oct             & HD 205478 & SB & 21 41 28.65  & -77 23 24.2    &  4.76    &   3.76 & K0   &  47.2   \\
$\delta$ Cap          & HD 207098 & Al & 21 47 02.45  & -16 07 38.2    &  3.16    &   2.87 & A7   &  84.6   \\
HD 216448             &           &    & 22 52 00.55  & +57 43 00.9    &  9.05    &   8.02 & K5   &  41.2   \\
$\gamma$ Tuc          & HD 219571 &    & 23 17 25.77  & -58 14 08.6    &  4.369   &   4.00 & F1   &  45.4   \\
$\lambda$ And         & HD 222107 & RS & 23 37 33.84  & +46 27 29.3    &  4.90    &   3.82 & G8   &  38.7   \\
\enddata
\end{deluxetable}

\clearpage

\begin{deluxetable}{llrrrrrcr}
\tabletypesize{\scriptsize}
\tablecolumns{9}
\tablewidth{0pc}
\tablecaption{Luminosity Class II Stars Within 30 pc}
\tablehead{ \colhead{Name}    &
\colhead{Alt. Name} &   \colhead{Type}     & \colhead{RA
J(2000)} & \colhead{Dec J(2000)}    & \colhead{B}   & \colhead{V}
& \colhead{Spec. Type} & \colhead{Parallax} }
\startdata

$\rho$ Pup            & HD 67523  & dS & 08 07 32.65  & -24 18 15.6    &  3.24  &   2.81 & F6  & 52.0       \\
R CrA                 & HIP 93449 & Or & 19 01 53.65  & -36 57 07.6    & 12.38  &  11.50 & A5  & 121.8      \\
\enddata
\end{deluxetable}


\begin{thebibliography}{}

\bibitem[]{146}
Allen, 1975 Astrophysical Quantities.

\bibitem[]{147}
Arrhenius S.A., 1903. Die Verbreitung des Lebens im Weltenraum.
Die Umschau 7: 481-485.
1980. The propagation of life in space, translated by Goldsmith D. In The Quest for
Extraterrestrial Life. Goldsmith D. (Ed.). University Science Books, Mill Valley, California,
32-33, 308 p.

\bibitem[]{147}
Becquerel P., 1910a, La Panspermie Interastrale devant les faits, Editions de
la revue politique et litt\'eraire (revue bleue) et de la revue
scientifique, Paris, 1910a, pp. 1-24.

\bibitem[]{147}
Becquerel P., 1910b, Comptes Rendus de l'Academie des Sciences,
Paris, 1910b, pp. 86-88.

\bibitem[]{148}
Beichman, C., Woolf, N.J., \& Lindensmith, C. 1999 (eds.),
The Terrestrial Planet Finder (TPF): A NASA Origins Program to
Search for Habitable Planets, JPL Publication 99-3.

\bibitem[]{152}
Brack A. 1993, 'Origine of Life', 3, 10.

\bibitem[]{153}
Brasier, M.D. et al. 2002, LPI, 33, 1614.

\bibitem[]{155}
Burchell M.J., Mann J., Bunch A.W., Brandao P.F.B., Icarus 2001, 154, 545.

\bibitem[]{155}
Charbonneau D., Noyes R. W., Korzennik S. G., Nisenson P., Jha S., 
Vogt S. S., Kibrick R. I. 1999, ApJ, 522, L145.

\bibitem[]{}
Chiosi, C., Bertelli, G., \& Bressan, A. 1992, ARAA, 30, 235. 

\bibitem[]{155}
Chyba C.F., Owen T.C. and Ip W.-H., 1994, 'Hazards due to Comets and Asteroids',
T. Gehrels Editor, The University of Arizona Press.

\bibitem[]{155}
Clampin M., Ford H.C., Illingworth G. and Petro L., 2001, AAS 199. 

\bibitem[]{}
Cox, A.N. 2000 "Allen's Astrophysical Quantities,'' 4th Edition, Edited by
A.N. Cox,  AIP Press (Springer, New York).

\bibitem[]{157}
Dalton, J.B. 2002, LPI, 33, 1555.

\bibitem[]{159}
Danchi W.C., Deming D., Kuchner M.J., Seager S. 2003a, ApJ, 597, L57.

\bibitem[]{159}
Danchi, W.C., et al. 2003b, in the Proceedings ``Towards 
Other Earths: Darwin/TPF and the Search for Extrasolar Terrestrial
Planets,''  Heidelberg, Germany, 22-25 April 2003, ESA Publication SP-539, p.83.

\bibitem[]{}
Despain, K. 1981, ApJ, 251, 639.

\bibitem[]{}
Deupree, R.G. 1996, ApJ, 471, 377.

\bibitem[]{160}
Forget, F., \& Pierrehumbert 1997, Sci, 278, 1273.

\bibitem[]{162}
Forget F., 1998, ``Earth, Moon and Planets'' 81, 59, Kluwer Academic Publishers.

\bibitem[]{164}
Fridlund, C.V.M., \& Gondoin, P. 2003, Proc. SPIE, 4852, 394.

\bibitem[]{165}
Frink S., Mitchell D.S., Quirrenbach A., Fischer D.A., Marcy G.W., Butler R.P. 2002,
ApJ 576, 478. 

\bibitem[]{167}
Furnes, H., Banarjee, N.R., Muelenbachs, K., Staudigel, H., 
de Wit M., 2004, Science, 304, 578.

\bibitem[]{170}
Gladman B.J., Burns J.A., Duncan M., Lee P. and Levinson H.F., 1996,
Science 271, 1387.

\bibitem[]{170}
Gough D.O., 1981, Solar Physics 74, 21.

\bibitem[]{170}
Holland H.D. 1997, Science 275, 38

\bibitem[]{170}
Iben I., 1967, ARAA 5, 571.

\bibitem[]{170}
Kasting J.F., Whitmire D.P. and Reynolds R.T., 1993, Icarus 101, 108.

\bibitem[]{170}
Kasting J.F., 1998, AAS 193, 5003.

\bibitem[]{172}
Kozlowski, M., Paczynski, B., 1975, Acta Astr, 25, 321.

\bibitem[]{173}
Maeder A. and Meynet G. 1988, AASS 76, 411

\bibitem[]{173}
Maher K.A., Stevenson D.J., 1988, Nature 331, 612.

\bibitem[]{173}
Mallik S.V., 1999, AA 352, 495.

\bibitem[]{173}
Mastrapa, R. M. E., Glanzberg, H., Head, J. N., Melosh, H. J., and
Nicholson, W. L., 2001, Earth
and Planetary Science Letters 189, 1.

\bibitem[]{173}
Meisel D.D., Diego Janches, John D. Mathews, 2002, ApJ 567, 323.

\bibitem[]{174}
Melosh, H.J., 2003, Astrobiology, 3, 207.

\bibitem[]{}
Mengel, J.G., \& Gross, P.G. 1975, Ap\&SS, 41, 407.  

\bibitem[]{177}
Mileikowsky C., 1997, in Astronomical and
Biochemical Origins and the
Search for Life in the Universe; Proceedings of the 5th International
Conference on Bioastronomy;
IAU Colloquium No. 161; Capri, July 1-5, 1996. Cosmovici C.B., Bowyer S.
and Werthimer D.
(Eds). Editrice Compositori, Bologna, 545-552, 814 p.

\bibitem[]{177}
Mileikowsky, C., Cucinotta, F. A., Wilson, J. W., Gladman, B., Horneck,
G., Lindegren, L.,
Melosh, J., Rickman, H., Valtonen, M., and Zheng, J. Q., 2000, Icarus 145,
391.

\bibitem[]{179}
Mischna, M.A. et al. 2000, Icar, 145, 546.

\bibitem[]{182}
Mojzsis, S.J. et al. 1997, Proc. SPIE, 3111, 162.

\bibitem[]{}
Neckel, H.  1975, A\&A, 42, 379.

\bibitem[]{184}
O'Keefe J.D. and Ahrens T.J., 1986, Science 234, 346.

\bibitem[]{}
Paczynski, B. 1970, AcA, 20, 47. 

\bibitem[]{}
Paczynski, B., \& Tremaine, S.D. 1977, ApJ, 216, 57.

\bibitem[]{184}
Portner D.M., Spiner D.R., Hoffman R.K. and Phillips C.R., 1961,
Science 134, 2047.

\bibitem[]{184}
Rasio F.A., Tout C.A., Lubow S.H. and Livio M., 1996, ApJ 470, 118.

\bibitem[]{184}
Rosing M.T., 1999, Science 283, 674.

\bibitem[]{184}
Roten C.-A. H., Gallusser A., Borruat G.D., UDRY S.D., Karamata D.,
Bulletin de la Soci\'et\'e Vaudoise des Sciences Naturelles
Vol 86 fasc.1 nov. 1998, pp. 1.

\bibitem[]{184}
Sackmann I.-J., Boothroyd A.I. and Kraemer K.E., 1993, ApJ 418, 457.

\bibitem[]{}
Scalo, J.M., Dominy, J.F., \& Pumphrey, W.A., 1978, ApJ, 221, 616. 

\bibitem[]{185}
Scalo, J.M., Miller, J.E., 1979, ApJ, 233, 569.

\bibitem[]{188}
Schopf J.W. 1993, Science, 260, 640.

\bibitem[]{188}
Schopf J.W., 1994, Proc. Natl. Acad. Sci. USA 91, 6735.

\bibitem[]{188}
Schr\"oder K.-P. and Sedlmayr E., 2001, 366, 913.

\bibitem[]{}
Schwarzschild, M. \& Harm, R.,  1962, ApJ, 136, 158. 

\bibitem[]{}
Harm, R., \& Schwarzschild, M.,  1964, ApJ, 139, 594.

\bibitem[]{190}
Sleep, N.H., Zahnle, K., and Neuhoff, P.S. 2001, PNAS, 98, 3666.

\bibitem[]{192}
Trauger J., et al. 2003, Proc. SPIE, 4854, 116.

\bibitem[]{193}
Weidemann J., 1987, AA 188, 74.

\bibitem[]{193}
Zubrin R., 2001, JBIS 54, 262.

\end{thebibliography}
\end{document}